\documentclass[pdflatex,sn-mathphys-num]{sn-jnl}

\usepackage{graphicx}%
\usepackage{multirow}%
\usepackage{amsmath,amssymb,amsfonts}%
\usepackage{amsthm}%
\usepackage{mathrsfs}%
\usepackage[title]{appendix}%
\usepackage{xcolor}%
\usepackage{textcomp}%
\usepackage{manyfoot}%
\usepackage{booktabs}%
\usepackage{algorithm}%
\usepackage{algorithmicx}%
\usepackage{algpseudocode}%
\usepackage{listings}%

\unnumbered

\begin{document}

\title[Uncertainty estimation and anomaly detection in chiral effective field theory studies of key nuclear electroweak processes]{Uncertainty estimation and anomaly detection in chiral effective field theory studies of key nuclear electroweak processes}

\author{\fnm{Bijaya} \sur{Acharya}}\email{bid@ornl.gov}

\affil{\orgdiv{Physics Division}, \orgname{Oak Ridge National Laboratory}, \orgaddress{\street{1 Bethel Valley Road}, \city{Oak Ridge}, \postcode{37831}, \state{TN}, \country{USA}}}

\abstract{Chiral effective field theory ($\chi$EFT) is a powerful tool for studying electroweak processes in nuclei. I discuss $\chi$EFT calculations of three key nuclear electroweak processes: primordial deuterium production, proton-proton fusion, and magnetic dipole excitations of $^{48}\mathrm{Ca}$. This article showcases $\chi$EFT's ability to quantify theory uncertainties at the appropriate level of rigor for addressing the different precision demands of these three processes.}


\maketitle

\section{Introduction}\label{sec:intro}

Chiral effective field theory ($\chi$EFT) is widely used to study nuclear structure and dynamics~\cite{Tews:2022yfb}. Based on the spontaneously broken approximate chiral symmetry of quantum chromodynamics, $\chi$EFT represents a low-energy expansion of the nuclear potential, formulated as a series in the ratio $Q/\Lambda$~\cite{Epelbaum:2008ga}. Here, $Q \sim\max(p,m_\pi)$, where $p$ denotes the typical momentum characteristic of the process under study and $m_\pi$ is the pion mass, while $\Lambda$ is the momentum scale beyond which the expansion breaks down. Operating within this framework, $\chi$EFT also generates nuclear electroweak currents~\cite{krebs2020} essential for describing nuclear $\beta,\gamma$ decays and nuclear responses to leptonic probes. $\chi$EFT thus not only enables precision few-body calculations of light nuclei~\cite{Piarulli:2017dwd,Maris:2020qne} but also provides reliable few-body ingredients that continue to drive progress in nuclear many-body computations in the era of {\it ab initio} modeling~\cite{Stroberg:2019bch,hu2022,Shen:2022bak}. 

One notable advantage of $\chi$EFT over traditional hadronic models~\cite{Reid:1968sq,Wiringa:1994wb,Machleidt:2000ge} lies in its capacity to systematically quantify theory uncertainties with varying levels of sophistication, tailored to the precision demand of the problem at hand. Examples of nuclear electroweak processes that motivate a detailed analysis of uncertainties are rates of nuclear electroweak reactions in astrophysical environments where theory inputs serve as proxies for unavailable or imprecise experimental data~\cite{Adelberger:1998qm,Adelberger:2010qa,Acharya:2016kfl}. Furthermore, the extraction of nuclear charge radii from precision spectroscopy of muonic atoms relies on the theoretical evaluation of the two-photon exchange contributions~\cite{Acharya:2020bxf,Ji:2013oba}. Similarly, neutrino experiments require a detailed theoretical understanding of how neutrinos interact with nuclei in the detector~\cite{Acharya:2019fij,Ruso:2022qes}. Such cases benefit from the utilization of $\chi$EFT predictions at various orders as ``data'', along with a statistical or machine-learning error model, to inductively infer a probability distribution for the truncation error of the $\chi$EFT expansion~\cite{Furnstahl:2014xsa,Furnstahl:2015rha}. Statistically interpretable truncation errors not only enable verification of assumptions made in the construction and application of $\chi$EFT but also facilitate the detection of statistical outliers---commonly referred to as ``anomalies'' in machine-learning literature---in the observed convergence pattern.

In this article, I focus on $\chi$EFT calculations for three key nuclear electroweak processes that are important in nuclear physics and astrophysics: the primordial deuterium production reaction, the proton-proton fusion reaction, and the magnetic dipole excitation of $^{48}\mathrm{Ca}$. The uncertainty analyses accompanying the theoretical predictions are detailed for the former two processes. The latter serves as an illustrative case, demonstrating how a simpler theory uncertainty analysis suffices to distinguish between two disparate experimental measurements. I refer the reader to Ref.~\cite{Epelbaum:2008ga} for a pedagogical review of nuclear EFTs and focus on specific applications in this article. 

\section{The primordial deuterium production reaction}\label{sec:bbn}

 The deuterium to hydrogen abundance ratio obtained from simulations of the big bang nucleosynthesis network~\cite{Serpico_2004,Pisanti:2007hk,Coc:2019rza,Pitrou:2018cgg,Gariazzo:2021iiu} aligns well with astronomical observations~\cite{Cooke:2017cwo,Planck:2018vyg}, serving as both a main evidence in favor of the big-bang theory and a test for extensions to the Standard Model. While modern simulations are based on experimental data for the deuterium burning reactions, they still use the theoretical result of Refs.~\cite{rupak,ando} for the production reaction, $np\rightarrow d\gamma$. The commonly adopted uncertainty of 0.2\% for this reaction comes from a simple estimate~\cite{rupak,ando} and is arbitrarily assigned a statistical meaning of ``$1\sigma$". A well-justified theory uncertainty estimate over the relevant [20~keV--200~keV] energy range, therefore, addresses a notable deficiency in an important physics problem. 
 A step in this direction was taken by Ref.~\cite{Acharya:2021lrv}, in which the error in the $np\rightarrow d\gamma$ cross section,  $\sigma_{np}(E)$, from truncation of the $\chi$EFT potential of Ref.~\cite{Reinert:2017usi} was estimated with fixed electromagnetic currents that included all $(Q/\Lambda)^{-2,-1,0}$ effects. 
 
\begin{figure}[h]
\centering
\includegraphics[width=\textwidth]{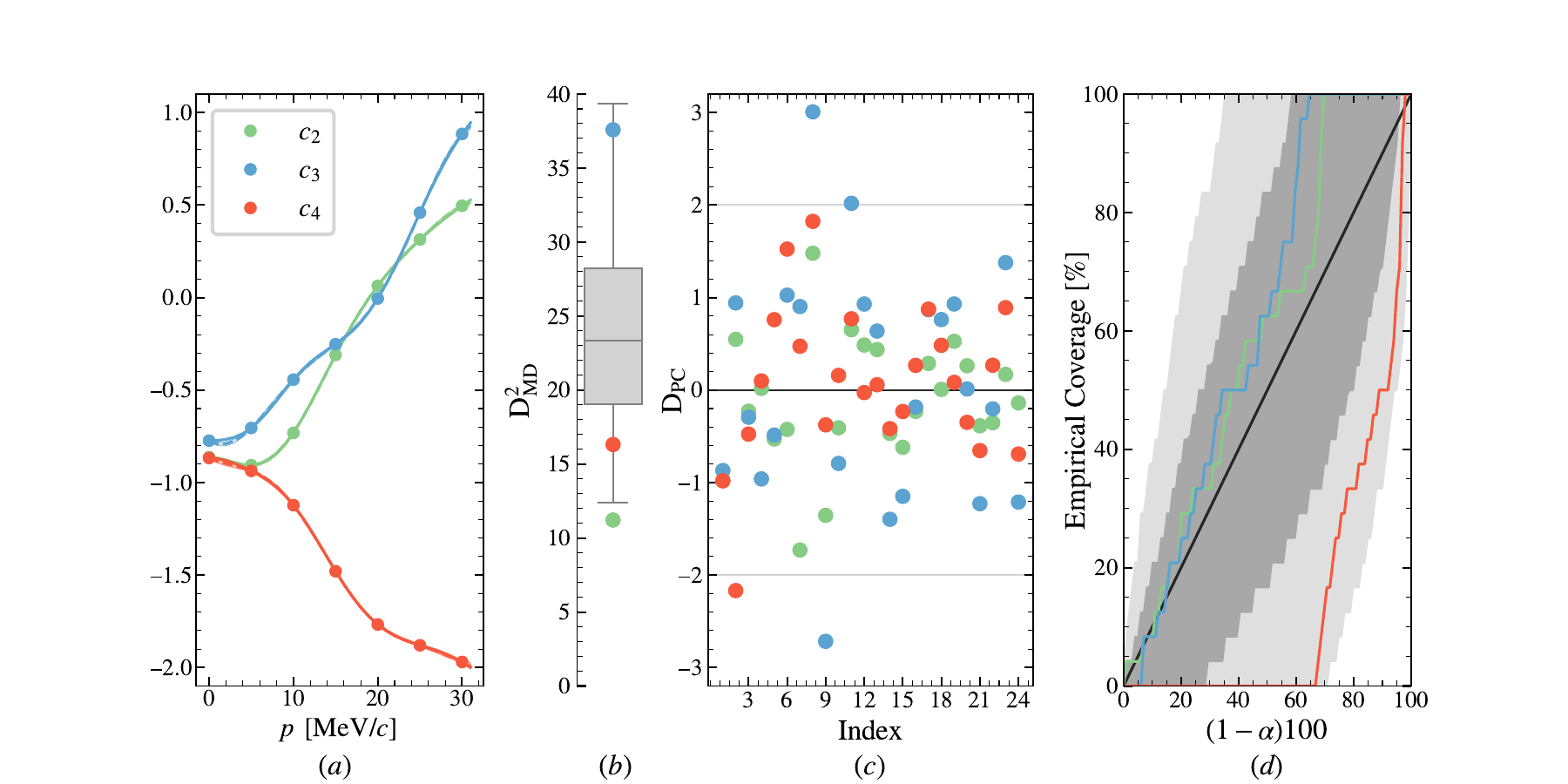}
\caption{The GP model of Ref.~\cite{Acharya:2021lrv} for analysis of the $\chi$EFT expansion coefficients and its diagnostic assessments. (a) Explicit $\chi$EFT calculations (solid lines) and corresponding GP emulations (dashed lines) with their respective $2\sigma$ intervals (bands). Filled circles represent training data, while validation data (four evenly spaced points between adjacent training points) are not shown. (b) The Mahalanobis distances relative to the mean (center line), with box plots indicating the $50\%$ credible interval and whiskers showing the $95\%$ credible intervals of the reference distribution. (c) The pivoted Cholesky diagnostics versus the index, accompanied by $95\%$ credible intervals represented by gray lines. (d) The credible-interval diagnostics, showcasing $1\sigma$ (dark gray) and $2\sigma$ (light gray) bands, calculated through sampling 1000 GP emulators. Created by Acharya and Bacca using the {\tt gsum}~\cite{Melendez:2019izc} package; originally published in  \url{https://doi.org/10.48550/arXiv.2109.13972}; licensed under CC BY 4.0.; reproduced with permission.}\label{fig:npdg_gp}
\end{figure}

\begin{figure}[h]
\centering
\includegraphics[width=0.8\textwidth]{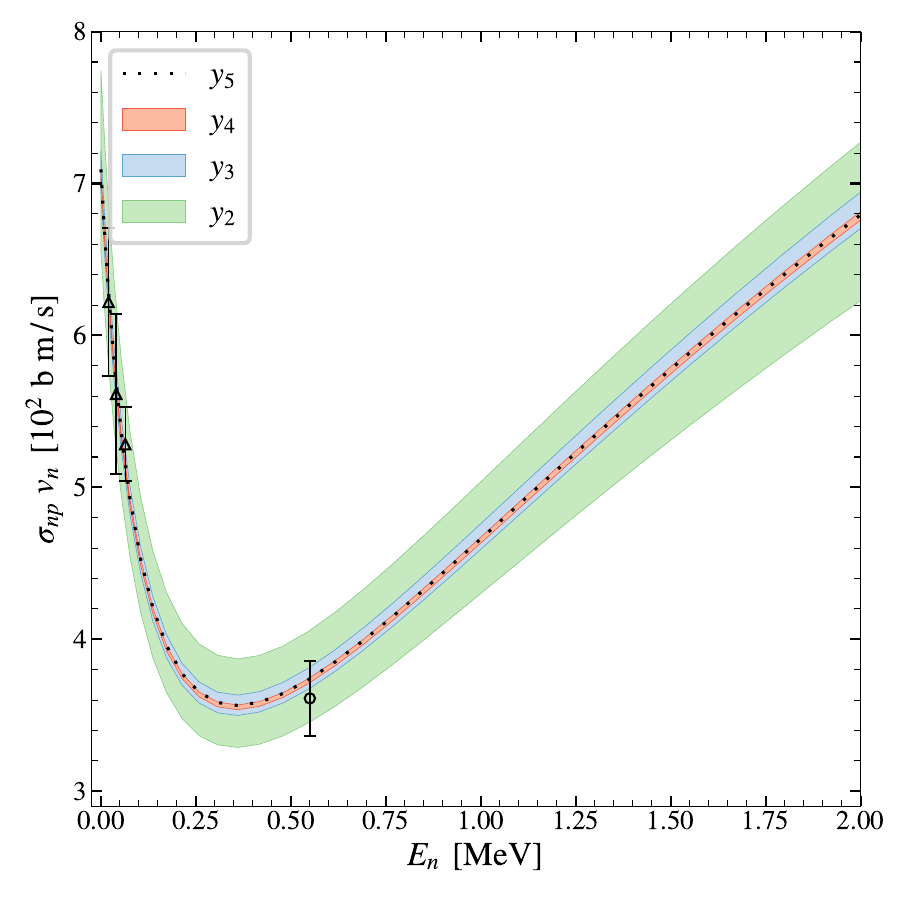}
\caption{The $2\sigma$ truncation error bands on the $\chi$EFT predictions $y_k$ at $k=2,3,4$ along with the prediction $y_5$ for $\sigma_{np} v_n$ is plotted versus the neutron energy $E_n$ in the rest frame of the proton. Theoretical results are from Ref.~\cite{Acharya:2021lrv} and experimental data are from Refs.~\cite{suzuki1995}~(triangles) and \cite{nagai1997}~(circle). Created by Acharya and Bacca; originally published in  \url{https://doi.org/10.48550/arXiv.2109.13972}; licensed under CC BY 4.0.; reproduced with permission.}\label{fig:npdg_pred}
\end{figure}

Let $v_n$ denote the neutron speed in the rest frame of the proton. The reaction rate is then proportional to  $\sigma_{np} v_n$. Ref.~\cite{Acharya:2021lrv} considered the order-$k$ $\chi$EFT prediction for this quantity, 
\begin{equation}
    y_k(p) = y_\mathrm{ref}(p) \sum_{n=0}^k c_n(p) \, [Q(p)/\Lambda]^n\,,
\end{equation}
and its EFT truncation error,  
\begin{equation}
\label{eq:trunc_err}
    \delta y_k(p) = y_\mathrm{ref}(p) \sum_{n=k+1}^\infty c_n(p) \, [Q(p)/\Lambda]^n\,.
\end{equation}
Here $y_k$ is a function of the $np$ relative momentum $p$, and  $y_\mathrm{ref}(p)$ is a dimensionful quantity such that the dimensionless coefficients $c_n(p)$ are smooth $\mathcal{O}(1)$ curves, provided that the EFT is converging expectedly. Following Ref.~\cite{Melendez:2019izc}, $c_n's$ were modeled as a Gaussian Process (GP). Various diagnostic assessments of the GP model are depicted in Fig.~\ref{fig:npdg_gp}. The GP model accurately emulates the full $\chi$EFT calculations and none of the analyzed orders shows noticeably anomalous behavior (see Ref.~\cite{Acharya:2021lrv} for details). Equipped with a validated GP model, Ref.~\cite{Acharya:2021lrv} obtained 95\% Bayesian credible intervals for $\delta y_k(p)$ under the assumption that they are uncorrelated across various values of $k$. The results, reproduced in Fig.~\ref{fig:npdg_pred}, show that the truncation errors for $k>2$ are much smaller than the experimental uncertainties of existing data. 

\section{The proton-proton fusion reaction}\label{sec:pp}

\begin{figure}[h]
\centering
\includegraphics[width=0.8\textwidth]{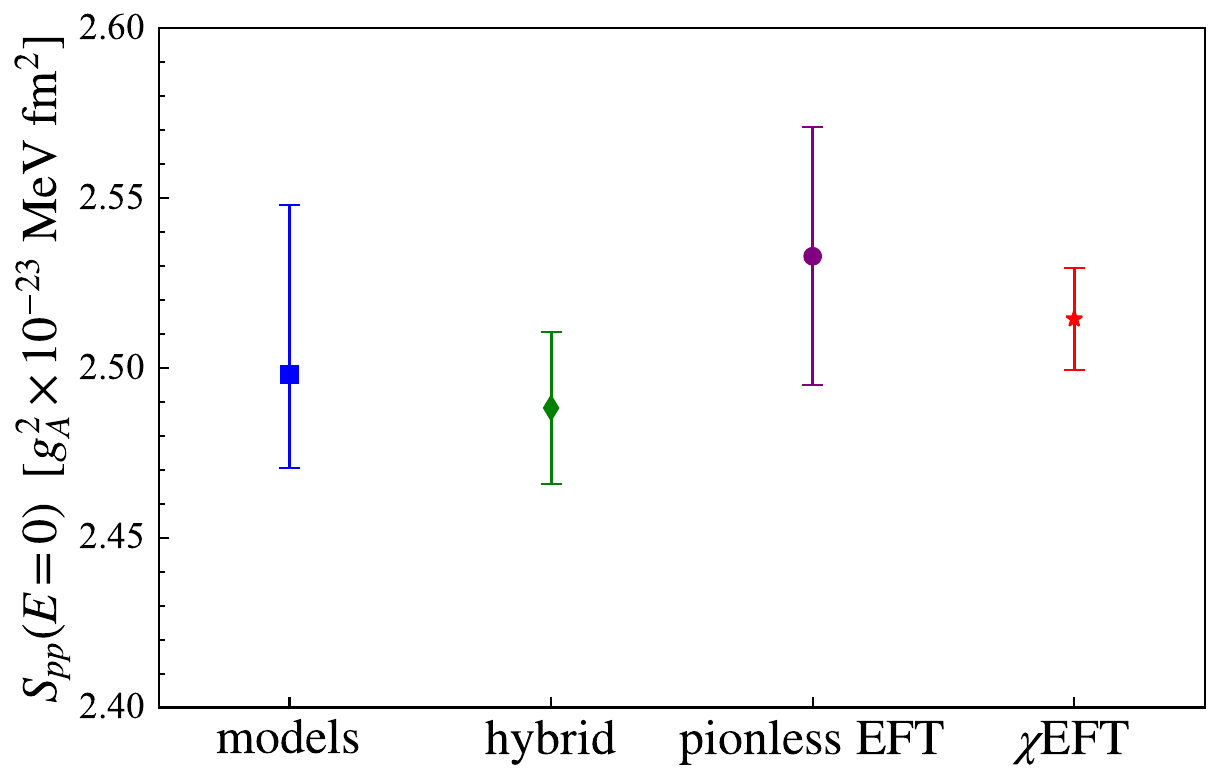}
\caption{The astrophysical $S$ factor for the proton-proton fusion reaction at zero energy $(E)$ in the units of $g_A^2\times10^{-23}\mathrm{MeV\,fm^2}$. The model-based recommended value from Ref.~\cite{Adelberger:1998qm}, the model/EFT-based hybrid recommended value from Ref.~\cite{Adelberger:2010qa} and the most recent pionless EFT result~\cite{De-Leon:2022omx} are compared to the $\chi$EFT prediction of Ref.~\cite{Acharya:2023xpd}. }\label{fig:spp}
\end{figure}

The rate of the reaction $p+p\rightarrow d+e^++\nu_e$, which sets the rate of hydrogen burning in light main-sequence stars such as the Sun, is too slow to measure in the laboratory. The astrophysical $S$ factor $S_{pp}(E)\equiv\sigma_{pp}(E)\,E\,\exp(2\pi\eta)$, where $\eta$ is the Sommerfeld parameter, has to be obtained from theory. Rigorous estimation of the accompanying uncertainty is important given ongoing progress in neutrino detection~\cite{Gonzalez-Garcia:2023kva} and the prospect of large-scale asteroseismic surveys~\cite{Rauer_2014}. The first $\chi$EFT calculation of this process was performed by Ref.~\cite{Marcucci:2013tda}, followed by Ref.~\cite{Acharya:2016kfl}, in which a detailed uncertainty analysis was performed using a family of $\chi$EFT interactions calibrated to several different pools of input data at several different regulator cutoffs. In the most recent and comprehensive $\chi$EFT study, Ref.~\cite{Acharya:2023xpd} compared prior calculations and obtained consistent values using four different $\chi$EFT interactions. Uncertainty analysis was performed for the $\chi$EFT interaction of Ref.~\cite{Reinert:2017usi} by estimating the truncation error as $(Q/\Lambda)^4$ times the maximum of $|c_0|$, $|c_2|$, or $|c_3|$, where $|c_k|$ are now the expansion coefficients for $S_{pp}(E=0)$ and are numerical values rather than functions of a kinematic variable. The result is compared with earlier recommended values and pionless EFT in Fig.~\ref{fig:spp}.

\section{The magnetic dipole excitation of $^{48}\mathrm{Ca}$}\label{sec:ca48}

\begin{figure}[h]
\centering
\includegraphics[width=0.8\textwidth]{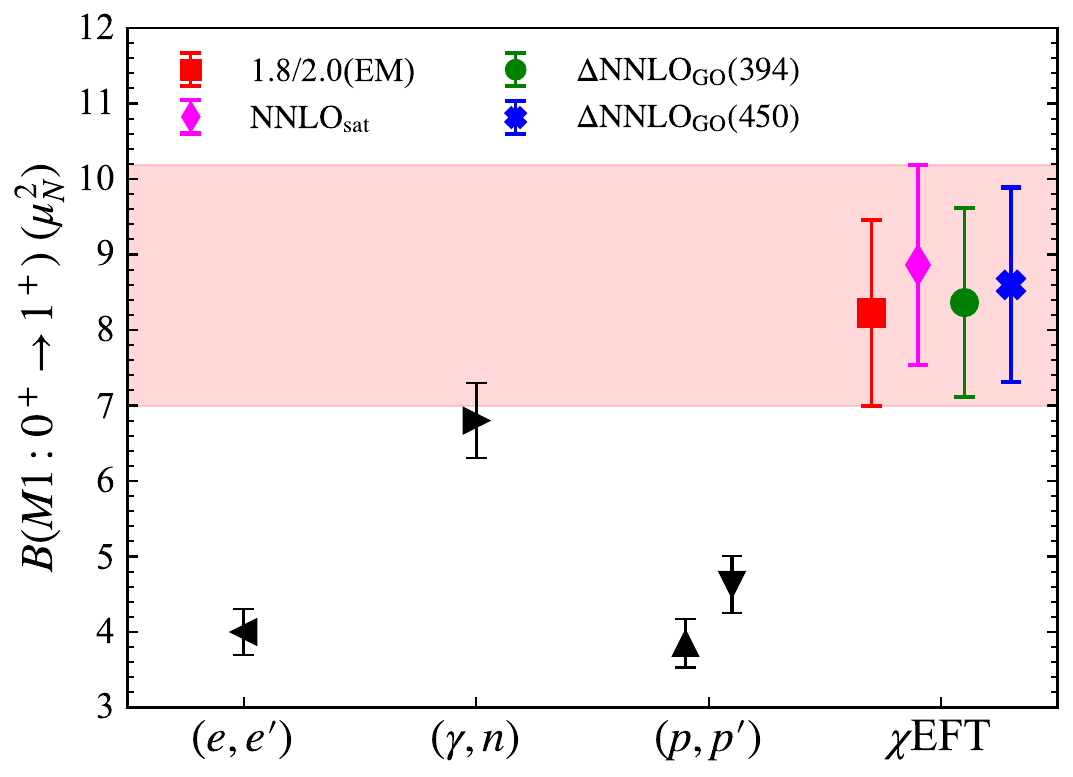}
\caption{The $B(M1)$ strength carried by the $1^+$ state at $10.23$~MeV in $^{48}$Ca. Data from the $(e,e^\prime)$~\cite{steffen1980}, $(\gamma,n)$~\cite{tompkins2011}, and $(p,p')$~\cite{birkhan2016} experiments are compared to $\chi$EFT predictions of Ref.~\cite{Acharya:2023ird} for four different $\chi$EFT interactions. Created by Acharya {\it et al}; originally published in  \url{https://doi.org/10.48550/arXiv.2311.11438}; licensed under CC BY 4.0.; reproduced with permission.}\label{fig:bm1}
\end{figure}

Magnetic dipole ($M1$) transitions are important in nuclear physics as they are clean probes of the spin-isospin structure of nuclei and are relevant in nuclear astrophysics due to their connection to weak processes. Particularly, the transition strength $B(M1)$ in $^{48}$Ca is crucial, influencing electron capture in nearby elements abundant in supernovae cores and affecting radiative neutron capture processes leading up to core collapse. $B(M1)$ distribution in $^{48}$Ca mainly centers on a single resonant state at $E_x=10.23$ MeV, with the reported values of $4\pm0.3~\mu_N^2$ for an $(e,e^\prime)$ experiment and $6.8\pm0.5~\mu_N^2$ for a $(\gamma,n)$ experiment. Indirect extractions based on $(p,p^\prime)$ experiments~\cite{birkhan2016} have claimed consistency with the $(e,e^\prime)$ measurement, which has circulated widely in the literature~\cite{Heyde:RevModPhys} and has been taken as evidence that the $M1$ strength is strongly quenched~\cite{birkhan2016,Heyde:RevModPhys}, analogous to the case of the Gamow-Teller (GT) strength. The two-body current contributions in $M1$ and GT processes are quite different in $\chi$EFT~\cite{krebs2020}. A comprehensive calculation accounting for wave function correlations, two-body currents, and continuum effects---considering the resonance's continuum nature---was performed using coupled-cluster theory~\cite{Hagen:2013nca} in Ref.~\cite{Acharya:2023ird}. A conservative 15\% uncertainty estimate was obtained by considering the sensitivity of the result to the $\chi$EFT interaction used, the precision of the many-body method, and the size of the leading two-body current effects included in the calculation. Here the theory precision provided by the leading order one- and two-body currents was sufficient to distinguish between the two experimental results, as shown in Fig.~\ref{fig:bm1}. The theoretical calculation favored the $(\gamma,n)$ experiment~\cite{tompkins2011}.

\section{Conclusion and outlook}\label{sec:summary}

In this work, I reviewed three recent applications of $\chi$EFT to nuclear electroweak processes: primordial deuterium production, proton-proton fusion, and magnetic dipole excitation of $^{48}\mathrm{Ca}$. I focused on uncertainty analysis, which was important for each of these processes because experiments were imprecise, unavailable, or in conflict. This work underscores how $\chi$EFT, powerful few- and many-body Schr{\"o}dinger equation solvers, and uncertainty quantification protocols come together to advance our understanding of nuclear phenomena across diverse contexts. Continued progress in these areas is critical for progress in nuclear physics. 

\bmhead{Acknowledgements}

I would like to thank the European Few-Body Research Committee and Few-Body Systems (Springer Nature) for conferring me with the Few-Body Systems Award 2022 for Young Professionals. I am grateful to my collaborators Sonia Bacca, Boris Carlson, Andreas Ekstr\"om, Christian Forss\`en, Gaute Hagen, Baishan Hu, Laura Marcucci, Petr Navr\'atil, Thomas Papenbrock, and Lucas Platter for their invaluable contribution to the work presented here. I am indebted to my PhD advisor Daniel Phillips whose mentoring has influenced my research trajectory that led to this research. This work is supported by the Office of High Energy Physics, Office of Science, U.S. Department of Energy, under Contract No. DE-AC02-07CH11359 through the Neutrino Theory Network Fellowship program.

\end{document}